\documentstyle[12pt,a4,epsfig]{article}

\newlength{\abstwidth}
\begin{document}
\newcommand{\hi }{{\footnotesize HIJING }}
\newcommand{\hij}{{\footnotesize HIJING}}

\thispagestyle{empty}
\begin{center}
{\Large {\bf Entropy and Multifractality in Relativistic Ion-Ion Collisions}}\\[5mm]

	{\bf Shaista Khan and Shakeel Ahmad\footnote{email: Shakeel.Ahmad@cern.ch\\Authors declare that there is no conflict of interest}}\\
{\it  Department of Physics, Aligarh Muslim University\\[-2mm] Aligarh - 202002 INDIA}

\end{center}

\begin{center}
{\bf Abstract}\\[2ex]
\begin{minipage}{\abstwidth}Entropy production in multiparticle systems is investigated by analysing the experimental data on ion-ion collision at AGS and SPS energies and comparing the findings with those reported earlier for hadron-hadron, hadron-nucleus and nucleus-nucleus collisions. It is observed that the entropy produced in limited and full phase space, when normalised to maximum rapidity exhibits a kind of scaling which is nicely supported by Monte Carlo model \hij. Using the R\'{e}nyi's order-q information entropy, multifractal characteristics of particle production are examined in terms of Generalized dimensions, D$_{q}$.  Nearly the same values of multifractal specific heat, \(c\) observed in hadronic and ion-ion collisions over a wide range of incident energies suggest that the quantity \(c\) may be used as a universal characteristic of multiparticle production in hadron-hadron, hadron-nucleus and nucleus-nucleus collisions.
 
{\footnotesize PACS numbers: 25.75--q, 25.75.Gz}\\[10ex]
\end{minipage}
\end{center}

%\begin{itemize}
\noindent KEY-WORDS: Entropy, Multifractality, Multiparticle production, Relativistic heavy-ion collisions.\\
%\end{itemize}
\phantom{dummy}

\newpage
\noindent {\bf 1. Introduction:}\\
\noindent Multiplicity distributions (MD) of particles produced in hadronic  and heavy-ion collisions play important role in extracting the first information on the particle production mechanism[1,2]. Investigations involving multiparticle production in hadron-hadron (hh), hadron-nucleus (hA) and nucleus-nucleus (AA) collisions at relativistic energies have been carried out by numerous groups during the last four decades[2,3,4,5,6]. However, the understanding of particle production mechanism still remains elusive. MD of relativistic charged particles produced in hh collisions have been observed to deviate from a Poisson distribution and are expected to provide information regarding the underlying production mechanism[4,7]. Asymptotic scaling of MD in hh collisions--the KNO scaling[8], predicted in 1972, was regarded as a useful phenomenological framework for comparing the MD at different energies ranging from $\sqrt{s}$ $\sim$ 10GeV to ISR energies[9]. It was, however pointed out[10,11] that KNO scaling is not strictly followed for inelastic hh collisions. This scaling law was observed to breakdown when collision energy reached to SPS range[4,7,9,12]. After the observations of KNO scaling violation in $\bar{p}p$ collisions at $\sqrt{s}$ = 540 GeV it was remarked that the observed scaling upto ISR energies was approximate and accidental[12]. A new empirical regularity, in place of KNO scaling, was then proposed[12] to predict the multiplicity distributions at different energies. It was shown that MD at different energies in full and limited rapidity ($\eta$) windows may be nicely reproduced by negative binomial distributions (N.B.D)[9,12,13]. These observations lead to revival of interest in studies involving MD and new scaling laws. Simak et al.[14], by introducing a new variable--the information entropy, showed that MD of charged particles produced in full and limited phase space in hh collisions exhibit a new type of scaling law in the energy range, $\sqrt{s}$ $\sim$ 19 to 900 GeV.\\

\noindent Analysis of the experimental data on $pp$, $\bar{p}p$ and $k^{+}p$ collisions over a wide energy range (upto $\sqrt{s}$ = 900 GeV) carried out by several workers[14,15,16] indicate that entropy increases with beam energy while the entropy per unit rapidity appears to be an energy independent quantity. These results indicate the presence of entropy scaling upto a few TeV energy. Presence of a similar scaling behavior in $pp$ collisions at LHC energies has also been reported by Mizoguchi and Biyajima[17] and S. Das et al.[2,4]. Analyses of AA collision data at AGS and SPS energies carried out by other  workers[1,6,18,19] too suggest that entropy produced in limited pseudorapidity ($\eta$) windows when normalized to the maximum rapidity is essentially independent of projectile and target mass as well as beam energy, indicating thereby the presence of entropy scaling.\\

\noindent Furthermore, the  generalization of R\'{e}nyi's order-q information entropy contains information on the multiplicity moments and can be used for investigating the multifractal characteristics of particle production[5,20,21]. It should be mentioned that this method of multifractal studies, is not related to the phase space bin-width or the detector resolution but to the collision energy[20]. It is, therefore considered worthwhile to carry out a well focused study of entropy production and subsequent scaling in AA collisions by analyzing the several sets of experimental data on AA collisions at AGS and SPS energies. R\'{e}nyi's information entropies are also estimated to investigate the multifractal characteristics of multiparticle production.\\

%=========================================================================================
%=========================================================================================
\noindent {\bf 2. Details of Data:}\\
 \noindent Six sets of events produced in collisions of $^{16}$O, $^{28}$Si, $^{32}$S and $^{197}$Au beams wih AgBr group of nuclei in emulsion at AGS and SPS energies, available in the laboratory are used in the present study. Details of these samples are presented in Table~1. These events are taken from the emulsion experiments performed by EMU01 Collaboration[22,23,24,25]. The other relevant details of the data, like criteria of selection of events, classification of tracks, extraction of AgBr group of interactions, method of measuring the emission angle, $\theta$ of relativistic charged particles, etc., may be found elsewhere[1,6,22,26,27,28].  \\

\noindent From the measured values of the emission angle $\theta$, the pseudorapidity variable, $\eta$, was calculated using the relation, $\eta$ = -ln$tan(\theta/2)$. It should be emphasized that the conventional emulsion technique has two main advantages over the other detectors: (i) its 4$\pi$ solid angle coverage and (ii) emulsion data are free from biases due to full phase space acceptance. In the case of other detectors, only a fraction of charged particles are recorded due to the limited acceptance cone. This not only reduces the multiplicity but may also distort some of the event characteristics like particle density fluctuations[6,26,29].
\noindent In order to compare the findings of the present work with the predictions of Monte Carlo model \hij [30,31], event samples corresponding to the experimental ones are simulated using the code \hij-1.35; the number of events in each simulated sample is equal to that in the corresponding real event sample. The events are simulated by taking into account the percentage of interactions which occur in the collision of projectile with various target nuclei in emulsion  constituting the AgBr group[22,25,26]. The value of impact parameter for each data sample was so set that the mean multiplicity of the relativistic charged particles becomes nearly equal to those obtained for the real data sets.\\

%=========================================================================================
%=========================================================================================
\noindent{\bf 3. Formalism}\\
\noindent Entropy of the charged particle multiplicity distribution, Shannon's information entropy is calculated using[14]\\
%\begin{equation}
\begin{eqnarray}
              S=-\sum_{n} P_{n}lnP_{n}
\end{eqnarray}
%\end{equation}
and its generalization, R\'{e}nyi's order q information entropy is estimated as[5,15,21]; \\
\begin{eqnarray}
          I_{q}=\frac{1}{q-1}ln\sum_{n}P_{n}^{q} \hspace{2ex} for \hspace{2ex} q \neq 1          
\end{eqnarray}

\noindent where, for q = 1  $\displaystyle{\lim_{q \to 1} I_{q} = I_{1} = S}$ 

\noindent and P$_{n}$ is the probability of production of n charged particles.
\noindent The generalized dimensions of order q may be estimated as[5,20,21] \\

\hspace{20ex} $D_{q} = I_{q}/Y_{m}$,  \nonumber \hspace{2ex}where  
\begin{eqnarray}
Y_{m} = ln[(\sqrt{s}-2m_{n}<n_{p}>)/m_{\pi}] = lnn_{max}
\end{eqnarray} 

\noindent Y$_{m}$ denotes the maximum rapidity in the centre-of-mass frame, $\sqrt{s}$ represents the center-of-mass energy, m$_{\pi}$ is the pion rest mass, $<n_{p}>$ denotes the average number of participating nucleons and n$_{max}$ is maximum multiplicity of relativistic charged particles produced for a given pair of colliding nuclei at a given energy.\\

%====================================================================================
%====================================================================================

%\newpage
\noindent{\bf 4. Results and discussion}\\
\noindent Probability P$_{n}$(${\Delta}{\eta}$) of production of n charged particles in a pseudorapidity window of fixed width is calculated by selecting a window of width ${\Delta}{\eta}$ = 0.5. This window is so chosen that its mid-position coincides with the center of symmetry of $\eta$ distribution, $\eta_{c}$. Thus, all the relativistic charged particles having their $\eta$ values lying in the range \((\eta_c-\Delta\eta/2) \leq \eta \leq (\eta_c+\Delta\eta/2)\) are counted to evaluate P$_{n}$. The window size is then increased in steps of 0.5 until the region, ${\eta}_{c}\pm0.3$ is covered. Values of entropy, S for different $\eta$-windows are calculated by using Eq.1, while the value of maximum rapidity is estimated from Eq.3. Variations of entropy normalized to maximum rapidity, S/Y$_{m}$ with $\eta$-window width, also normalized to  maximum rapidity, ${\Delta}{\eta}/Y_{m}$ for the  experimental and \hi data sets are plotted in Fig.1. It is observed in the figure that S/Y$_{m}$  first increases upto ${\Delta}{\eta}/Y_{m}\sim 0.5$ and thereafter acquires almost a constant value. It is interesting to note that the data points for various samples of events overlap to form a single curve. This indicates the presence of entropy scaling in AA collisions at AGS and SPS energies. Results from \hi simulated events also support the presence of entropy scaling.\\

\noindent Similar entropy scaling in AA collisions has been reported by us[1,6] and also by the other workers[18,21] for central and minimum bias events. In our earlier work, attempt was made to ensure whether the observed entropy scaling is of dynamical nature. For this purpose the correlation free Monte Carlo events samples (Mixed events) corresponding to each of the real data samples were generated and analyzed. Those findings reveal that the entropy scaling observed is the distinct feature of the data and is of dynamical origin[1,6].\\

\noindent According to Eq.3 the quantity, $(\sqrt{s}-2m_{n}<n_{p}>)/m_{\pi}$ is equal to the maximum charged particle multiplicity at a given center of mass energy. It would, therefore, be convenient to examine the mean multiplicity \(<n_{s}>\) vs entropy in limited $\eta$ windows as well as in full $\eta$ range; the entropy in the entire $\eta$ range, S$_{max}$ is calculated using Eq.1. Variations of S$_{max}$ with lnE$_{total}$ for the experimental and \hi events are exhibited in Fig.2: E$_{total}$ denotes the total energy of the beam nucleus. It may be noticed in the figure that S$_{max}$ increases linearly with lnE$_{total}$ for both real as well as simulated event samples. \hij, however, predicts somewhat smaller values of S$_{max}$ as compared to the corresponding S$_{max}$ values estimated from the real data. These findings are, thus, in agreement with those reported for $^{16}$O-nucleus collision at 60A and 200A GeV/c and $^{32}$S-nucleus interaction at 200 GeV/c. Shown in Fig.3 are the variations of S/S$_{max}$ with \(<n>/<n>_{max}\) for the six data sets considered. It is evident from the figure that the data points corresponding to various types of collisions almost overlap to form a single curve. It may also be noted that S/S$_{max}$ $\rightarrow$ 1.0 as \(<n>/<n>_{max}\) $\rightarrow$ 1.0. This observations too support the presence of entropy scaling in AA collisions at AGS and SPS energies.\\

\noindent It has been shown[5,32] that the constant specific heat, widely used in standard thermodynamics, reflects multifractal character of various stochastic systems in a reasonable approximation. Such a constancy in multifractal specific heat has been observed in hadronic and ion-ion collisions[33,34] by analysing the data adopting the method proposed by Takagi[35]. As mentioned earlier, the R\'{e}nyi's order-q information entropy may also be used to understand the multifractal nature of particle production and estimation of multifractal specific heat. The advantage of this method is that it does not depend on the phase space bin width and hence on detector resolution, rather it accounts for the fractal resolution which is related to the collision energy[5]. From the definitions of R\'{e}nyi's information entropy, I, and generalized dimensions, D$_{q}$ (Eqs.1,2,3), it is evident that for a given q, $(I_{q})_{max}$ = ln$n_{max}$. The highest entropy is achieved for the greatest 'chaos' of a uniformly distributed probability function $P_{n} = 1/n_{max}$[21] and Eq.2 gives $D_{q} = I_{q}/(I_{q})_{max}$. \\

\noindent Variations of D$_{q}$ with q for various event samples are shown in Fig.4. It is observed that $D_{q}$ monotonously decreases with increasing order q and the trend of decrease for the real and \hi events are nearly the same except that \hi predicts somewhat smaller values of $D_{q}$ as compared to that for experimental data. The $D_{q}$ spectrum for order q $\ge$ 2, which for multifractals is decreasing function of q, can be related to the scaling behaviour of q point correlation integrals[5,36]. Thus, the observed trend of $D_{q}$ against q observed in the present study indicates the multifractal nature of multiplicity distributions in full phase space in ion-ion collisions at the energies considered. The presence of multifractality, although predicts $D_{q}$ to decrease with q, yet no further useful information about the q-dependence of $D_{q}$ spectrum can be extracted from which conclusions about the scaling properties of q-correlation integrals[5] may be drawn. It has been suggested[5,32] that in constant heat approximation, $D_{q}$ dependence on q acquires the following simple form:
 
 \begin{eqnarray}
D_{q} \simeq (a-c)+c\frac{lnq}{q-1}
\end{eqnarray} 

\noindent where a is the information dimension, $D_{1}$, while c denotes the multifractal specific heat. The linear trend of variation $D_{q}$ with lnq/q-1, given by Eq.4, is expected to be observed for multifractals. On the basis of classical analogy with specific heat of gases and solids the value of c is predicted[37] to remain independent of temperature in a wide range of q. In order to test the validity of Eq.4, $D_{q}$ values are plotted against lnq/q-1 in Fig.5. It is evident from the figure that $D_{q}$ increases linearly with lnq/q-1. The lines in the figure are due to the best fits to the data obtained using Eq.4. The values of the parameters 'a' and 'c', thus, obtained are listed in Table~2. It is interesting to note from the table that the values of multifractal specific heat, c for all the data sets, are nearly the same, $\sim$ 0.2 indicating its independence of the beam energy and mass. It may also be noticed from the table that the experimental values of  c are quite close to those predicted by \hi. It is worthwhile to mention that the values of multifractal specific heat obtained in the present study are close to those obtained by us[34] by analysing some of these data sets using Takagi's approach[35]. Incidentally similar values of c has been reported by Bershadskii[38] for 10.6A GeV/c $^{197}$Au-nucleus collisions. In p-nucleus interactions too, the value of multifractal specific heat has been observed to be $\sim$ 0.25 in the energy range, 200-800 GeV[21,32,38]. \\

\noindent These findings, thus, indicate that the constant-specific approximation is applicable to the multiparticle production in relativistic hadronic and ion-ion collisions. Moreover, nearly the same values of multifractal specific heat, c observed in the present study as well as by the other workers using the data on hh, hA and AA collisions involving various projectiles/targets at widely different energies does indicate that the parameter c may be regarded as a universal characteristics of hadronic and heavy-ion collisions.\\

%===========================================================================
%===========================================================================

\noindent{\bf 5. Summary}\\
\noindent On the basis of the findings of the present work the following conclusions may be arrived at:
\begin{enumerate} 
\item[1.] The entropy normalized to maximum rapidity exhibits a saturation beyond ${\Delta}{\eta}/Y_{m}$ $\sim$ 0.5 indicating thereby the presence of large amount of entropy around mid-rapidity region.
\item[2.] Overlapping data points for various sets of events in S/Y$_{m}$ vs ${\Delta}{\eta}/Y_{m}$ plots exhibit entropy scaling in AA collisions at AGS and SPS energies.
\item[3.] The scaling observed with the experimental data is nicely supported by \hi model.
\item[4.] The decreasing values of $D_{q}$ with increasing order number q may be taken as a signal of multifractal nature of multiplicity distributions of relativistic charged particle produced in hadronic and heavy-ion collisions.
\item[5.] Besides the information dimension, $D_{1}$, there is yet another parameter, c which may be used as a universal parameter for multiparticle production in high energy hadronic and heavy-ion collisions.\\
\end{enumerate} 
%=========================================================================
%=========================================================================

\noindent  {{\bf References}}
\begin{enumerate}
\item[1] Shakeel Ahmad et al., Int. J. Mod. Phys.E22 (2013) 1350088.
\item[2] S. Das et al., Nucl. Phys.A862 (2011) 438.
\item[3] R. M. Weiner, Int. J. Mod. Phys.E15 (2006) 37.
\item[4] S. Das et al., arXiv: 1104.3053v1[hep-ph], (2011).
\item[5] M. K. Suleymanov et al., arXiv: 0304206[hep-ph], (2003).
\item[6] Shakeel Ahmad et al., Adv. in High En. Phys.2013 (2013) 836071.
\item[7] G. J. Alner et al., (UA5 Collab.), Phys. Lett.B138 (1984) 304.
\item[8] Z. Koba, N. B. Nielsen and P.Olesen, Nucl. Phys.B40 (1972) 317.
\item[9] G. J. Alner et al., (UA5 Collab.) ,Phys. Lett.B167 (1986) 476.
\item[10] A. Wroblewski et al., Acta Phys. Polon.B4 (1973) 857.
\item[11] W. Thome et al., Nucl. Phys.B129 (1977) 365.
\item[12] G. J. Alner et al., (UA5 Collab.) Phys. Lett.B160 (1985) 199.
\item[13] G. J. Alner et al., (UA5 Collab.) Phys. Lett.B160 (1985) 193.
\item[14] V. Simak et al., Phys. Lett.B206 (1988) 159.
\item[15] P. A. Carruthers et al., Phys. Lett.B212 (1988) 369.
\item[16] M. R. Ataian et al., (ENS/NA22 Collab.) Acta Phys. Polon.B36 (2005) 2969.
\item[17] T. Mizoguchi and M. Biyajima, Euro Phys. J.C70 (2010) 1061.
\item[18] D. Ghosh et al., J. Phys.G38 (2011) 065105.
\item[19]  D. Ghosh et al., Euro Phys. Lett.65 (2004) 311.
\item[20] M. Pachr et al., Mod. Phys. Lett.A7 (1992) 2333.
\item[21] A. Mukhopadhyay et al., Phys. Rev.C47 (1993) 410.
\item[22] M. I. Adamovich et al., (EMU01 Collab.) J. Phys.G22 (1996) 1469.
\item[23] M. I. Adamovich et al., (EMU01 Collab.) Phys. Lett. B227 (1989) 285.
\item[24] M. I. Adamovich et al., (EMU01 Collab.) Phys. Rev. Lett. 65 (1990) 412.
\item[25] M. I. Adamovich et al., (EMU01 Collab.) Phys. Lett.B201 (1988) 397.
\item[26] Shakeel Ahmad et al., Phys. Scr.87 (2013) 045201.
\item[27] Shakeel Ahmad et al., J. Phys.G30 (2004) 1145.
\item[28] Shakeel Ahmad et al., Int. J. Mod. Phys., E24 (2015) 1550074.
\item[29] M. L. Cherry et al., Acta Phys. Polon. B29 (1998) 2129.
\item[30] X. N. Wang, Phys. Rep. 280 (1997) 287.
\item[31] X. N. Wang and M. Gyulassy, Comp. Phys. Comm.83 (1994) 307.
\item[32] A. Bershadskii, Physica A253 (1998) 23.
\item[33] D. Ghosh et al., Phys. Rev.C65 (2002) 067902.
\item[34] Shakeel Ahmad et al., Chaos Solitons and Fractals, 42 (2009) 538. 
\item[35] F. Takagi, Phys. Rev. Lett.72 (1994) 32.
\item[36] G. Paladin and A. Vulpiani, Phys. Rep. 156 (1987) 147.
\item[37] L. D. Landau, E. M. Lifschitz, {\it Statistical Physics, Part I} (Pergamon Press, Oxford, 1980).
\item[38] A. Bershadskii, EuroPhys. J.A4 (1999) 283.		
\end{enumerate}

%================= Table-1 =============================================

\newpage
\noindent Table~1: Number of events selected for the analysis.
\begin{center}
\begin{tabular}{|c|c|c|} \hline
Energy (GeV/c) & Type of Interactions & No. of Events   \\ [2mm] \hline
%(GeV/c) & interactions &  Events   \\[2mm] \hline

10.6A    &     $^{197}$Au-AgBr	&  577 \\ [2mm]  \hline
14.5A	 &       $^{16}$O-AgBr     &  379 \\ [2mm]  \hline
14.5A     &       $^{28}$Si-AgBr    &  561 \\[2mm]  \hline
60A      &       $^{16}$O-AgBr     &  422 \\[2mm]  \hline
200A      &     $^{16}$O-AgBr     &  223 \\[2mm]  \hline
200A     &       $^{32}$S-AgBr     & 452 \\[2mm] \hline 
       
\end{tabular}
\end{center}

%========= Table 2 ==============================
\newpage
\noindent Table~2: Values of parameters a and c, occurring in Eq.4 for the various event samples.
\begin{center}
\begin{tabular}{|c|c|c|c|c|c|} \hline
Type of & & \multicolumn{2}{|c|}{Expt.} & \multicolumn{2}{|c|}{HIJING} \\ \cline{3-6}
interactions & Energy (GeV/c)&   a &  c    &  a &  c \\[2mm] \hline
 $^{197}$Au-AgBr & 10.6A  &  0.85 $\pm$ 0.06  &  0.20 $\pm$ 0.06  &  0.78 $\pm$ 0.06  &  0.22 $\pm$ 0.06 \\[2mm] \hline
 $^{16}$O-AgBr   & 14.5A  &  0.86 $\pm$ 0.08  &  0.22 $\pm$ 0.08  &  0.90 $\pm$ 0.09  &  0.21 $\pm$ 0.09 \\[2mm] \hline
 $^{28}$Si-AgBr  & 14.5A  &  0.91 $\pm$ 0.07  &  0.20 $\pm$ 0.07  &  0.85 $\pm$ 0.06  &  0.17 $\pm$ 0.06 \\[2mm] \hline
 $^{16}$O-AgBr   & 60A    &  0.92 $\pm$ 0.07  &  0.22 $\pm$ 0.07  &  0.90 $\pm$ 0.08  &  0.23 $\pm$ 0.08 \\[2mm] \hline
 $^{16}$O-AgBr   & 200A   &  0.82 $\pm$ 0.08  &  0.20 $\pm$ 0.07  &  0.80 $\pm$ 0.08  &  0.21 $\pm$ 0.08 \\[2mm] \hline
 $^{32}$S-AgBr   & 200A   &  0.89 $\pm$ 0.06  &  0.21 $\pm$ 0.06  &  0.88 $\pm$ 0.06  &  0.20 $\pm$ 0.06 \\[2mm] \hline
              
\end{tabular}
\end{center}

%======================================================================
\newpage
\begin{figure}[]
\begin{center}\mbox{\psfig{file=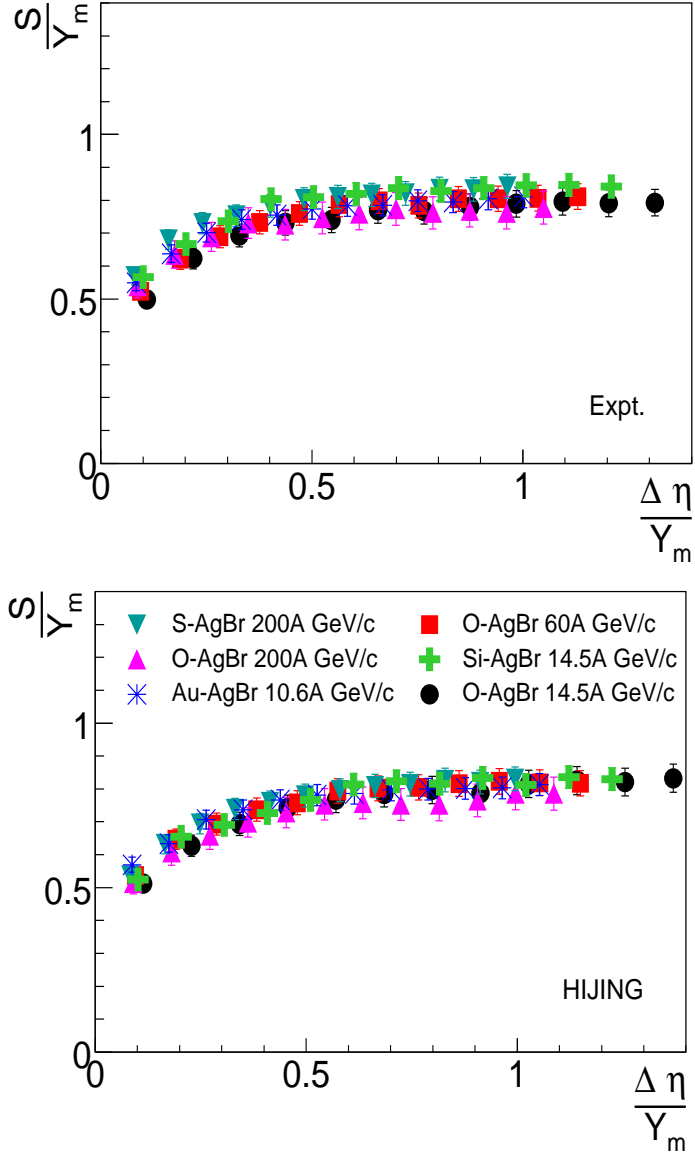,width=10cm,height=16cm}}
\end{center}
\caption[Fig1.]{\sf{Variations of entropy normalized to maximum rapidity, S/Y$_{m}$ with $\eta$-window width normalized to  maximum rapidity, ${\Delta}{\eta}/Y_{m}$ for the Experimental (top) and \hi data (bottom)}}
\label{lindat}
\end{figure}

\newpage
\begin{figure}[]
\begin{center}\mbox{\psfig{file=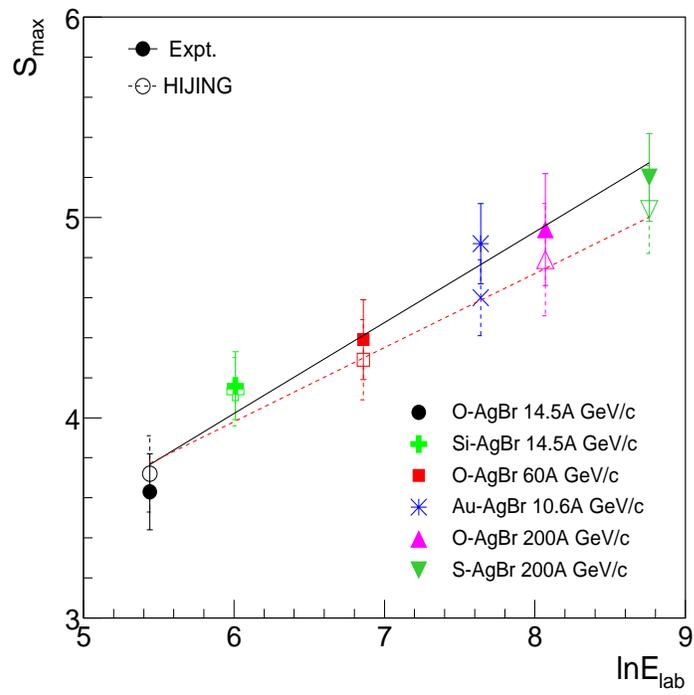,width=10cm,height=10cm}}
\end{center}
\caption[Fig1.]{\sf{Variations of S$_{max}$ with lnE$_{total}$ for the experimental and \hi events.}}
\label{lindat}
\end{figure} 

\newpage
\begin{figure}[]
\begin{center}\mbox{\psfig{file=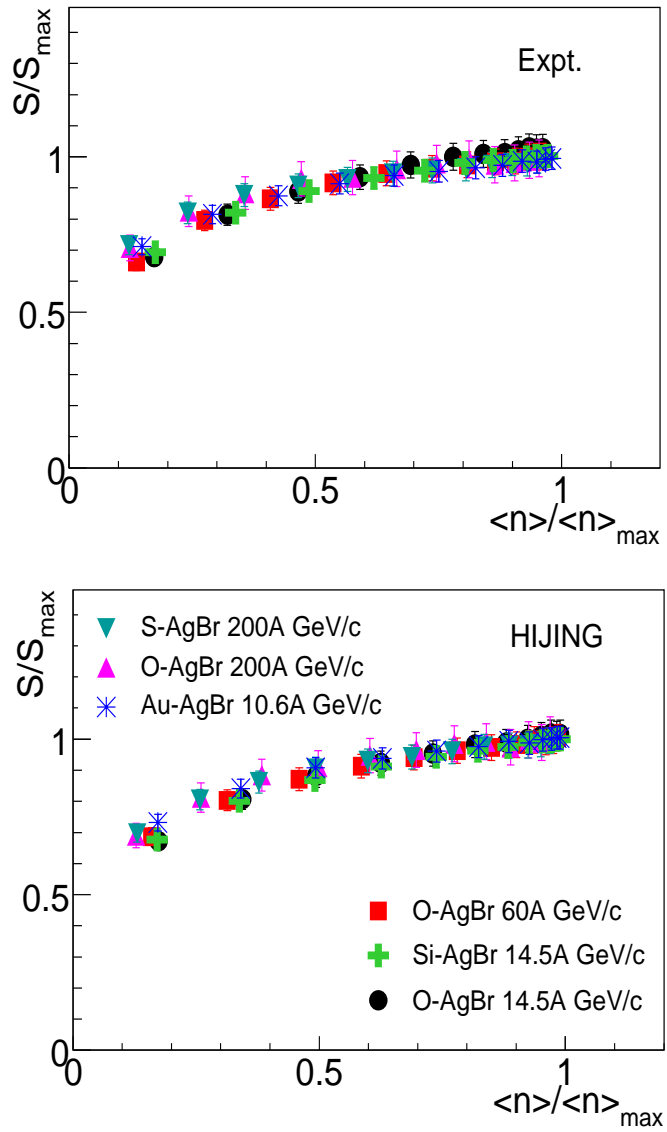,width=10cm,height=16cm}}\end{center}
\caption{\sf Variations of S/S$_{max}$ with \(<n>/<n>_{max}\) for the experimental and \hi event samples.}
\label{lindat}
\end{figure}

\newpage
\begin{figure}[]
\begin{center}\mbox{\psfig{file=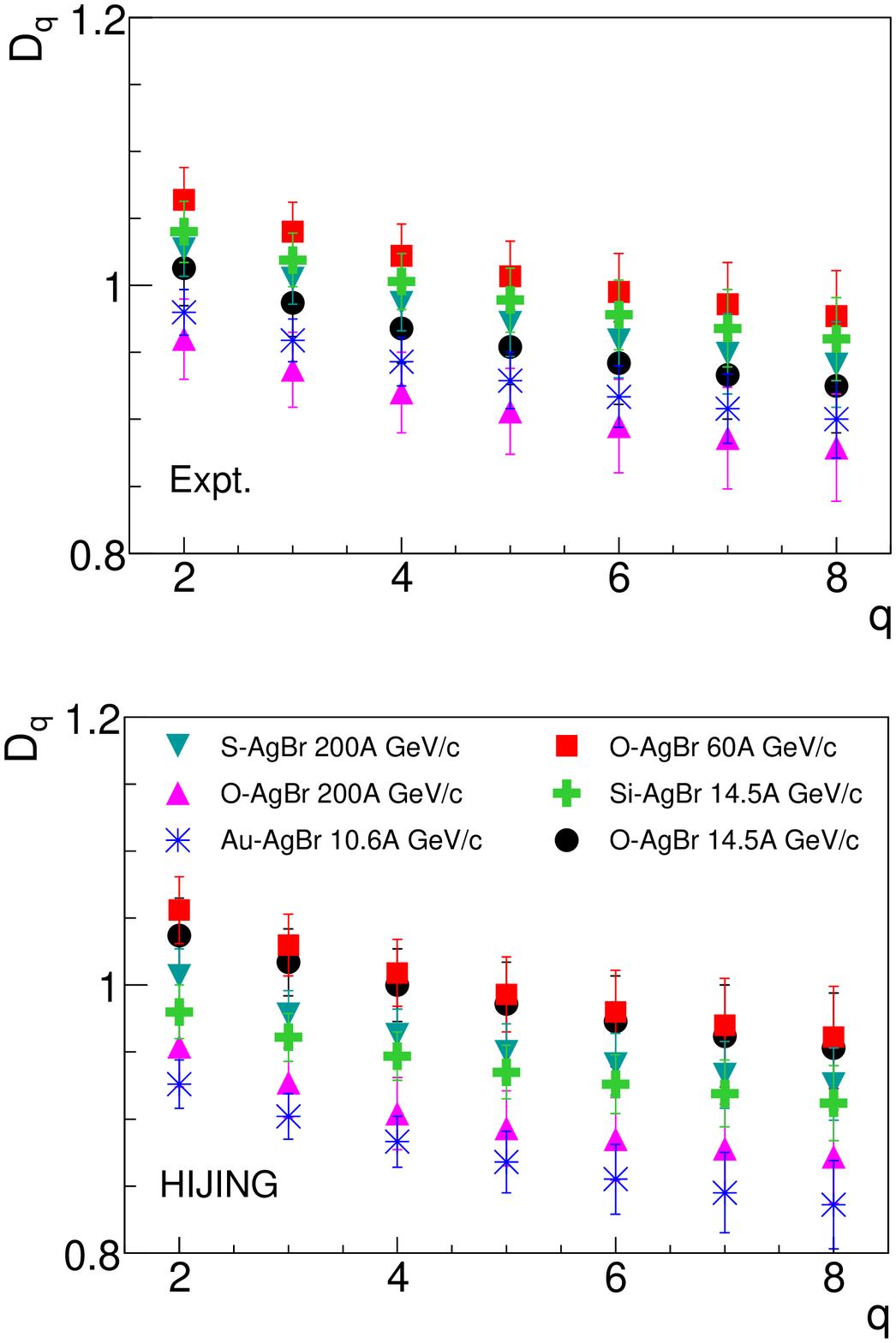,width=10cm,height=16cm}}\end{center}
\caption{\sf Dependence of D$_{q}$ on q for  experimental and \hi events. }
\label{lindat}
\end{figure}

\newpage
\begin{figure}[hptb]
\begin{center}\mbox{\psfig{file=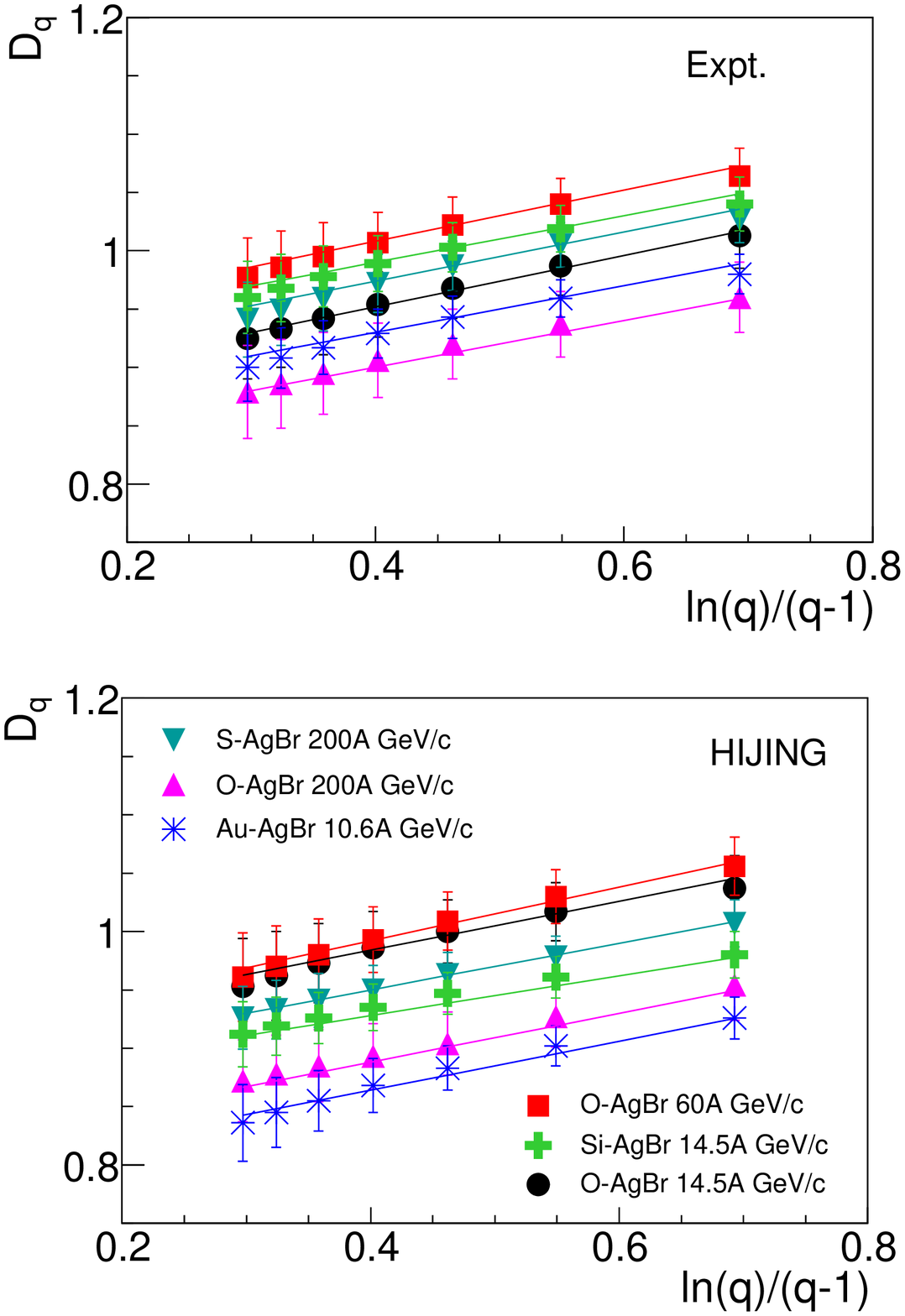,width=10cm,height=16cm}}\end{center}
\caption{{\sf D$_{q}$ vs ln(q)/(q-1) plots for the experimental and \hi data at various energies.}}
\label{lindat}
\end{figure}

\end{document}